\documentclass[aps,prl,reprint,showpacs,superscriptaddress]{revtex4-1}
\usepackage{amsmath}
\usepackage{graphicx}
\usepackage{bm}
\begin{document}
\title{Probing Quantitatively Two-electron Entanglement with a Spintronic Quantum Eraser}
\author{Wei Chen}
\affiliation{National Laboratory of Solid State Microstructures and Department of Physics, Nanjing University, Nanjing 210093, China}
\affiliation{Department of Physics and Center of Theoretical and Computational Physics, The University of Hong Kong, Pokfulam Road, Hong Kong, China}
\author{R. Shen}
\email{shen@nju.edu.cn}
\affiliation{National Laboratory of Solid State Microstructures and Department of Physics, Nanjing University, Nanjing 210093, China}
\author{Z. D. Wang}
\email{zwang@hku.hk}
\affiliation{Department of Physics and Center of Theoretical and Computational Physics, The University of Hong Kong, Pokfulam Road, Hong Kong, China}
\author{L. Sheng}
\affiliation{National Laboratory of Solid State Microstructures and Department of Physics, Nanjing University, Nanjing 210093, China}
\author{B. G. Wang}
\affiliation{National Laboratory of Solid State Microstructures and Department of Physics, Nanjing University, Nanjing 210093, China}
\author{D. Y. Xing}
\affiliation{National Laboratory of Solid State Microstructures and Department of Physics, Nanjing University, Nanjing 210093, China}
\begin{abstract}
We design an ingenious spintronic quantum eraser to quantitatively probe the two-electron entanglement. It is shown that the concurrence of two spin-entangled electrons is directly given by the Aharonov-Bohm oscillation amplitude of the Fano factor, a measurable current-current correlation, making it rather promising to experimentally quantify the two-electron entanglement. The singlet and triplet entangled states are distinguished by the opposite signs in the Fano factor. Since the main building blocks in the designed setup, an electronic Mach-Zehnder interferometer and a spin filter, have already been implemented, our proposal is particularly pertinent to experiments.
\end{abstract}
\maketitle
Entanglement of electron pairs in solids is a key resource for large-scale implementation of quantum information and computation schemes. Recently, the generation of  spin-entangled electrons via Cooper pair splitting has been theoretically proposed \cite{Lesovik} and experimentally observed \cite{Hofstetter}. Nevertheless, one central question is still very challenging: how to detect whether the two electrons are spin-entangled and to what extent they are entangled.

One way to demonstrate the entanglement is Bell tests \cite{Bell,Clauser}. The maximal violation of the Bell inequality can give the concurrence \cite{Gisin}, which is a measure of the two-particle entanglement \cite{Wootters}. Several proposals for Bell tests in solid-state devices have been theoretically suggested \cite{Kawabata}. In order to find the maximal violation caused by two entangled spins, one needs to precisely measure the correlations between two spins with arbitrary polarization directions, which is a high demand in solid-state experiments.

On the other hand, the quantum eraser consisting of two entangled photons has already been realized experimentally in quantum optics \cite{Zeilinger}. In principle, there are two steps to achieve a quantum optics eraser \cite{Scully}. First, the which-way information (WWI) of the signal photon is registered and, according to the  complementarity principle, the decoherence of the signal photon occurs. Second, a proper measurement is performed on the entangled partner of the signal photon to erase the WWI so that the interference recurs.

\begin{figure}
\centering
\includegraphics[width=0.48\textwidth]{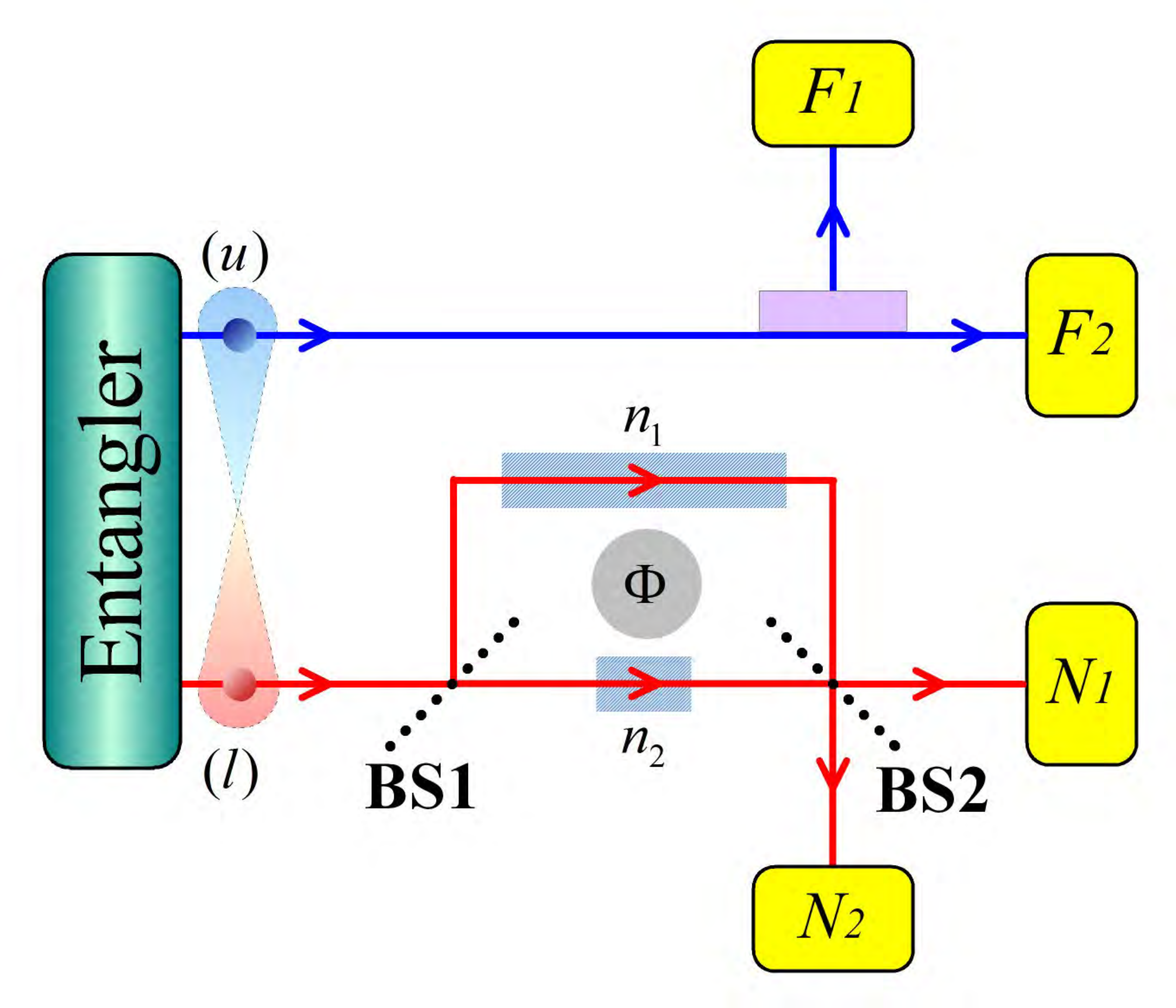}
\caption{(color). Illustration of a spintronic quantum eraser. Two entangled electrons are injected from the entangler into channels \textit{u} and \textit{l}, respectively. The electron in channel \textit{u} travels through a spin filter (shaded area) and reaches leads $F_{1}$ and $F_{2}$, respectively. The electron in channel \textit{l} travels through an MZI and reaches leads $N_{1}$ and $N_{2}$, respectively.  The MZI consists of two beam splitters, BS1 and BS2, labeled by dotted lines. The Rashba spin-orbit coupling exists in the shaded regions in paths $n_{1}$ and $n_{2}$, and a magnetic flux $\Phi$ is enclosed by the two paths.} \label{fig1}
\end{figure}

In this Letter, we show that two entangled electrons can form a spintronic quantum eraser, and more importantly, the eraser serves as a better substitute for the Bell test to quantitatively probe the two-electron entanglement. It is found that the Fano factor exhibits the Aharonov-Bohm oscillation with its amplitude being simply a quadratic function of the concurrence.

The proposed quantum eraser is sketched in Fig. \ref{fig1}.The entangler represents a sink of spin-entangled electrons. On the theoretical side, an entangler can be realized by use of superconductors \cite{Lesovik} or quantum dots \cite{Loss}. On the experimental side, with a bias under the superconducting gap, the spin-entangled electrons can be efficiently created via Cooper pair splitting \cite{Hofstetter}. The two electrons from the entangler are separately injected into two channels along $x$ direction, labeled by \textit{u} and \textit{l}, respectively. Each channel consists of only a single transverse mode. We denote the electron in channel \textit{u} as the idler electron (IE) and that in channel \textit{l} as the signal electron (SE). A longer decoherence length at low temperatures is beneficial to the observation of the entanglement so that we restrict our discussions at zero temperature for simplicity. The state of two spin-entangled electrons is expressed by
\begin{equation}\label{wave0}
\big|\Psi_{0}\big\rangle=\Big[\sqrt{\kappa}a^{\dagger}_{l\uparrow}(E_{1})a^{\dagger}_{u\downarrow}(E_{2})
\mp\sqrt{1-\kappa}a^{\dagger}_{l\downarrow}(E_{1})a^{\dagger}_{u\uparrow}(E_{2})\Big]\big|0\big\rangle,
\end{equation}
where operators $a^{\dagger}_{i\sigma}(E)$ create particles with energy $E$ and spin $\sigma$ in channel $i$ ($i=u,l$), and satisfy the discrete anticommutation relations $\{a_{i\sigma}(E_{1}),a^{\dagger}_{i'\sigma '}(E_{2})\}=\delta_{ii'}\delta_{\sigma\sigma'}\delta_{E_{1}E_{2}}$. The vacuum state $|0\rangle$ is the filled Fermi sea. The concurrence \cite{Wootters} of the entangled state Eq. (\ref{wave0}) is given by $\mathcal{C}=2\sqrt{\kappa(1-\kappa)}$. Particularly, $\kappa =0,1$ ($\mathcal{C}=0$) and $\kappa =1/2$ ($\mathcal{C}=1$) represent a direct-product state and a maximally entangled state, respectively, while other values of $\kappa$  correspond to general entangled states with $0<\mathcal{C}<1$. For maximally entangled states, the minus and plus in Eq. (\ref{wave0}) represent the singlet and triplet states, respectively.

$|\Psi_{0}\rangle$ in Eq. (\ref{wave0}) is the initial state of two electrons emitted from the entangler. The final state after scattering can be obtained by investigating all the transport processes of the SE and the IE, respectively. The SE in channel \textit{l} travels through a Mach-Zehnder interferometer (MZI) and finally reaches leads $N_{1}$ and $N_{2}$, respectively. An MZI consists of two beam splitters (BS1 and BS2) and a magnetic flux $\Phi$ enclosed by two paths, as shown in Fig. \ref{fig1}, which has already been realized for electrons in experiments \cite{Ji}.

There is no backscattering when the SE walks through BS1 \cite{Ji,Liu}, and the transmission amplitudes into paths $n_{1}$ and $n_{2}$ are assumed to be $t_{1}$ and $t_{2}$, respectively. The wave function of the electrons after BS1 evolves to
\begin{equation}\label{wave1}
\begin{split}
\big|\Psi_{1}\big\rangle=\Big[&\sqrt{\kappa}\big(t_{1}a^{\dagger}_{n_{1}\uparrow} +t_{2}a^{\dagger}_{n_{2}\uparrow}\big)a^{\dagger}_{u\downarrow}\\
&\mp\sqrt{1-\kappa}\big(t_{1}a^{\dagger}_{n_{1}\downarrow}+t_{2}a^{\dagger}_{n_{2}\downarrow}\big) a^{\dagger}_{u\uparrow}\Big]\big|0\big\rangle,
\end{split}
\end{equation}
where the energy indexes, $E_{1}$ for the SE and $E_{2}$ for the IE, are omitted for simplicity. One finds that whether spin-up or spin-down SE travels through both paths. The interference pattern between two paths is determined by the relative phase $\varphi =\mathrm{Arg}(t_{1}t_{2}^{*})$. Since the relative phase is controlled by the magnetic flux enclosed by two paths, the interference effect here is the so-called AB oscillation \cite{Gefen}. The BS1 in solids can be prepared by the quantum point contact in two-dimensional electron gas, and the strongest interference occurs with $|t_{1}|=|t_{2}|=1/\sqrt{2}$, which is realizable in experiments \cite{Ji,Liu}.

However, in our spintronic MZI as shown in Fig. \ref{fig1}, a finite Rashba spin-orbit coupling $\alpha_{R}$ \cite {Rashba} exists in the Rashba regions in two paths, which substantially alters the interference pattern. The Rashba Hamiltonian for electrons moving along $x$ direction one-dimensionally is given by $H_{R}=-\alpha_{R}k_{x}\sigma_{y}$, where $k_{x}$ is the wave vector and $\sigma_{y}$ is the Pauli matrix \cite{Rashba}. The transfer of the SE through the Rashba region can be described by a unitary operator $U=\exp(-i\theta_{R}\sigma_{y}/2)$ \cite{Loss2}, implying a spin rotation about the $y$ axis by an angle $\theta_{R}=-2m^{*}\alpha_{R}L/\hbar^{2}$, where $m^{*}$ is the effective mass of the electron and $L$ is the length of the Rashba region. We set the length of the Rashba region in path $n_{1}$ to be three times of that in path $n_{2}$ so that the spin rotation angles in the two paths satisfy $\theta_{n_{1}}=3\theta_{n_{2}}$. By tuning gate voltages, $\alpha_{R}$ and therefore the spin rotation angles can be controlled. We adopt here $\theta_{n_{1}}=3\theta_{n_{2}}=-3\pi /2$, which can be achieved in realistic systems. For example, the parameters in InGaAs/InAlAs systems are $m^{*}=0.046m_{0}$ with $m_{0}$ being the mass of a free electron and $\alpha_{R}\approx 0.39\times10^{-11}$ eV$\cdot$m \cite{Datta}, and thus a spin rotation angle of $-3\pi /2$ can be obtained with a Rashba region of the length about 1 $\mu$m.

After the Rashba spin precession, the wave function of the electrons can be derived by applying transfer operator $U$, and is written as
\begin{equation}\label{wave2}
\begin{split}
\big|\Psi_{2}\big\rangle=\Big[&\sqrt{\kappa}\big(-t_{1}a^{\dagger}_{n_{1}x} +t_{2}a^{\dagger}_{n_{2}\bar{x}}\big)a^{\dagger}_{u\downarrow}\\
&\mp\sqrt{1-\kappa}\big(t_{1}a^{\dagger}_{n_{1}\bar{x}}+t_{2}a^{\dagger}_{n_{2}x}\big) a^{\dagger}_{u\uparrow}\Big]\big|0\big\rangle,
\end{split}
\end{equation}
where subscripts $x$ and $\bar{x}$ denote the spin-$x$ and spin-$\bar{x}$ states, two orthogonal states with spins polarized parallel and antiparallel to the $x$ direction, respectively. One finds that the WWI of the SE is registered by both electron spins. For example, if the SE is spin-$x$ and the IE is spin-down, the SE must travel in path $n_{1}$. In other words, the traveling paths of the SE are completely distinguishable by use of both electron spins. According to the complementarity principle, the vanishing of the AB oscillation should be expected. We note that, as shown in Eq. (\ref{wave1}), the IE alone can not destroy the interference between two paths. It is the Rashba spin-orbit coupling here that acts as a WWI detector. Our scheme is a spintronic version of the quantum eraser.

The scattering at BS2 is similar to that at BS1 and can be described by a spin-independent unitary matrix $S=\begin{pmatrix}
r&t\\t'&r'
\end{pmatrix}$, where amplitudes $r$ and $t$ correspond to transfers from path $n_{1}$ and path $n_{2}$ to lead $N_{1}$, respectively, and amplitudes $t'$ and $r'$ correspond to transfers into lead $N_{2}$. By incorporating matrix $S$ into Eq. (\ref{wave2}), the electron state after BS2 reads
\begin{equation}\label{wave3}
\begin{split}
\big|\Psi_{3}\big\rangle &=\Big[\sqrt{\kappa}\big(-t_{1}rb^{\dagger}_{N_{1}x}-t_{1}t'b^{\dagger}_{N_{2}x} +t_{2}tb^{\dagger}_{N_{1}\bar{x}}
\\&+t_{2}r'b^{\dagger}_{N_{2}\bar{x}}\big)
a^{\dagger}_{u\downarrow}\mp\sqrt{1-\kappa}\big(t_{1}rb^{\dagger}_{N_{1}\bar{x}}+ t_{1}t'b^{\dagger}_{N_{2}\bar{x}}
\\&+t_{2}tb^{\dagger}_{N_{1}x}+t_{2}r'b^{\dagger}_{N_{2}x}\big)
a^{\dagger}_{u\uparrow}\Big]\big|0\big\rangle,
\end{split}
\end{equation}
where operators $b^{\dagger}_{j\sigma}$ create spin-$\sigma$ electrons in lead $j$, representing the outgoing waves after scattering.

In order to find the final state of the two entangled electrons, one should also consider the transportation of the IE, which travels through a spin filter and reaches leads $F_{1}$ and $F_{2}$, respectively. There are no restrictions for electrons going into lead $F_{2}$, but only spin-$y$ electrons (spins polarized along $y$ direction) can penetrate through the spin filter (shaded region in Fig. \ref{fig1}) and arrive at lead $F_{1}$. In other words, the spin-$\bar{y}$ electrons are excluded from lead $F_{1}$ and can only reach lead $F_{2}$, and the spin-$y$ electrons can reach either leads $F_{1}$ or $F_{2}$. Such a spin filter can be achieved experimentally with a half-metal tunnel junction \cite{Moodera} or a Zeeman-split quantum dot \cite{Hanson} attached between channel $u$ and lead $F_{1}$. Both spin filters of half-metal junctions and quantum dots can generate an output current nearly 100\% spin-polarized \cite{Moodera, Hanson}.

The scattering for the IE at the spin filter can be described by the following relations
\begin{equation}\label{substitution}
\begin{split}
&a_{u\uparrow}^{\dagger}=\frac{1}{\sqrt{2}}\gamma_{1}b^{\dagger}_{F_{1}}+\frac{1}{\sqrt{2}}\gamma_{2}b^{\dagger}_{F_{2}y}+\frac{1}{\sqrt{2}}b^{\dagger}_{F_{2}\bar{y}},\\
&a_{u\downarrow}^{\dagger}=\frac{-i}{\sqrt{2}}\gamma_{1}b^{\dagger}_{F_{1}}+\frac{-i}{\sqrt{2}}\gamma_{2}b^{\dagger}_{F_{2}y}-\frac{-i}{\sqrt{2}}b^{\dagger}_{F_{2}\bar{y}},
\end{split}
\end{equation}
where $\gamma_{1,2}$ are amplitudes for spin-$y$ electrons transmitted into leads $F_{1,2}$, respectively. The final state of the two entangled electrons is given by inserting Eq. (\ref{substitution}) into Eq. (\ref{wave3}).

The current contributed by a pair of entangled electrons emitted from the entangler can be obtained by the standard scattering matrix approach. The current operator in lead $j$ ($j=N_{1},F_{1}$) is given by \cite{Burkard}
\begin{equation}\label{current}
I_{j}(t)=\frac{e}{h\nu}\sum_{EE'\sigma}b^{\dagger}_{j\sigma}(E)b_{j\sigma}(E')\exp\big[i(E-E')t/\hbar\big],
\end{equation}
where $\nu$ is the density of states at the Fermi level.

The currents in leads $N_{1}$ and $F_{1}$ are obtained by the expectation values of $I_{j}(t)$ in the final state. One finds that $\langle I_{F_{1}}\rangle=(e/h\nu )(\Gamma_{1}/2)$ and $\langle I_{N_{1}}\rangle=(e/h\nu )(T_{1}R+T_{2}T)$, with $\Gamma_{1}=|\gamma_{1}|^{2}$, $T_{1,2}=|t_{1,2}|^{2}$, $R=|r|^{2}$, and $T=|t|^{2}$. Both currents are constants determined only by the transmission probabilities. Since there is a finite Rashba spin-orbit coupling in paths $n_{1}$ and $n_{2}$ acting as a WWI detector, the interference between two paths is completely destroyed. The current in lead $N_{1}$ can be understood in a straightforward manner. The SE arrives at lead $N_{1}$ through two paths. One is to move along path $n_{1}$ and then from path $n_{1}$ to lead $N_{1}$, resulting in a current $T_{1}R$. The other is to move along path $n_{2}$ and then from path $n_{2}$ to lead $N_{1}$, resulting in a current $T_{2}T$. The total current in lead $N_{1}$ is just a sum of these two contributions.

In order to recover the AB oscillation, the WWI must be erased by a proper measurement on the IE. If the SE and the IE are not entangled, the measurement on the IE could not recover the interference, and the maximal recovery will occur if two electrons are maximally entangled. One can expect that the amount of the recovered interference will give the information on the measure of entanglement, concurrence.

In quantum optics, the interference recurs with a coincidence counting of the correlated measurements on entangled photons.  The counterpart here is the shot noise, a current-current correlation, between leads $N_{1}$ and $F_{1}$. In general, the noise power between leads $j$ and $j'$ is given by \cite{Burkard}
\begin{equation}\label{noise}
S_{jj'}(\omega )=\lim_{\tau\rightarrow\infty}\frac{h\nu}{\tau}\int^{\tau}_{0}dt e^{i\omega t}\langle\delta I_{j}(t)\delta I_{j'}(0)\rangle,
\end{equation}
where $\delta I_{j}=I_{j}-\langle I_{j}\rangle$, and the average is made under the final state. With the help of Eqs. (\ref{wave3})-(\ref{current}), the zero-frequency ($\omega =0$) noise power between leads $N_{1}$ and $F_{1}$ is found to be
\begin{equation}\label{noise_N1F1}
S_{N_{1}F_{1}}=\pm\frac{e^{2}}{h\nu}\mathcal{C}\Gamma_{1}\sqrt{T_{1}T_{2}RT}\sin\varphi,
\end{equation}
where the phase difference between the two scattering amplitudes at BS2, $\mathrm{Arg}(rt^{*})$, is incorporated into the definition of $\varphi$. It is seen clearly that the AB oscillation recurs in the shot noise. Although all the transmission probabilities in Eq. (\ref{noise_N1F1}) can be obtained beforehand, the strongest interference pattern occurs with $T_{1}=T_{2}=1/2$ and $R=T=1/2$, which can be experimentally realized for beam splitters \cite{Ji,Liu}. Under this condition, the Fano factor, defined as $F=(S_{N_{1}F_{1}}/\langle I_{F_{1}}\rangle )/(e/2)$, is simply expressed by
\begin{equation}\label{Fano}
F=\pm\mathcal{C}\sin\varphi .
\end{equation}

Here come our main results. First, the Fano factor in Eq. (\ref{Fano}) is a sinusoidal function of the phase controlled by the external magnetic flux, revealing the recovery of the AB oscillation and a quantum eraser effect in solids. Second, the two-electron entanglement is quantitatively probed by the Fano factor. The concurrence is just the amplitude of the AB oscillation. Third, the Fano factor in a spintronic quantum eraser can also distinguish the singlet and triplet states. The different signs in Eq. (\ref{Fano}) represent a $\pi$-phase shift in the AB oscillation. For the maximally entangled states ($\mathcal{C}=1$), the plus and the minus are corresponding to the singlet and triplet states, respectively.

Notably, although Eq. (\ref{Fano}) is obtained with an initial state spin-polarized along $z$ direction, the present scenario is still valid in a more general case. Given the initial state with spin  $\sigma=\cos\alpha\sigma_{z}+\sin\alpha\sin\beta\sigma_{y}+\sin\alpha\cos\beta\sigma_{x}$, with $\alpha$ and $\beta$ being two arbitrary polarization angles, the noise power can be obtained as $S_{N_{1}F_{1}}=(e^{2}\Gamma_{1} /4h\nu)[\mathcal{C}^{2}\sin^{2}\alpha\sin^{2}\beta\pm \mathcal{C}(\cos^{2}\beta+\sin^{2}\beta\cos^{2}\alpha)]\sin\varphi$. For $\alpha=\beta=0$, Eq. (\ref{noise_N1F1}) is recovered. 

More practically, the entangled electrons emitted from a superconductor are spin-unpolarized and the measured noise power is an average over angles $\alpha$ and $\beta$. The average current in lead $F_{1}$ remains unchanged and the Fano factor is given by $F=(1/3)(\mathcal{C}^{2}\pm 2\mathcal{C})\sin\varphi$. In this case, the Fano factor still shows an AB oscillation behavior with an amplitude being a quadratic function of the concurrence. Considering an experimentally observed AB oscillation $F=\eta A\sin\varphi$ ($A>0$), the concurrence is obtained as
\begin{equation}\label{concurrence}
\mathcal{C}=\eta \left(\sqrt{1+3\eta A}-1\right),
\end{equation}
where $\eta=\pm 1$ correspond to the entangled pairs from the singlet and triplet superconductors, respectively.

We here wish to pinpoint that the present spintronic quantum eraser is distinctly different from some other schemes of mesoscopic erasers \cite{eraser}. Our proposal consists of two entangled electrons and utilizes the spin-orbit coupling as the WWI detector. More importantly, the concurrence can be quantitatively determined from the AB oscillation of the Fano factor, which serves as a better substitute for the Bell test and paves a new but direct way for probing quantitatively the two-electron entanglement.

It is also worthwhile to compare the present method for entanglement detection with the Bell test. Both methods can demonstrate the entanglement, however, there are at least two important differences between them. First,  the working principles are different. The present method employs an idea of quantum eraser, which is a result of the complementarity principle instead of the Bell's theorem. Second, with respect to quantifying the concurrence of two spin-entangled electrons, the present method is simpler than the Bell test. In a Bell test, the concurrence is given by the maximal violation of the Bell inequality \cite{Gisin}. To achieve this goal, one needs to precisely measure the spin-spin correlations, noting that a perfect separation and counting of spins along arbitrary polarization directions is a high demand in experiments. In the present method, the concurrence can simply be determined from the AB oscillation of the Fano factor, which is a charge current correlation. Although we use a spin filter in channel $u$, a perfect separation and counting of different spins is not required. Since we do not require $\Gamma_{1}=1$, the spin-$y$ and spin-$\bar{y}$ electrons in channel $u$ are not fully separated. The only prerequisite is that the output current in lead $F_{1}$ is fully spin-polarized, which is realizable in experiments \cite{Moodera, Hanson}.

In order to find the concurrence of two spin-entangled electrons with the present method, two simple steps are sufficient. First, the gate voltages and subsequently the Rashba spin-orbit coupling are tuned so that the AB oscillation of the current in lead $N_{1}$ is eliminated. Second, the concurrence is evaluated by measuring the AB oscillation of the Fano factor.

The main building blocks of the proposed eraser are an MZI and a spin filter, both of which have already been implemented \cite{Ji, Moodera, Hanson}. In addition, the order of magnitude of the decoherence length in metal and semiconductor mesoscopic systems is usually several micrometers at low temperatures \cite{Lin}. For GaAs, the spin decoherence length can even exceed 100 $\mu$m \cite{Kikkawa}. Therefore, the spin decoherence length in a mesoscopic system is long enough to realize the proposed device of a scale about 1 $\mu$m. In this sense, our proposal is very likely to be realized experimentally in the near future.

In summary, we have designed a spintronic quantum eraser to quantitatively probe the two-electron entanglement. It has been found that the concurrence is simply determined from the amplitude of the AB oscillation of the Fano factor. Our proposal paves a new but direct way for probing quantitatively the two-electron entanglement, which may be regarded as a better substitute for the corresponding Bell test.

\begin{acknowledgments}
We would like to thank Yang Yu and Zheng-Yuan Xue for fruitful discussions. This work was supported by 973 Program (Grants No. 2011CB922103, No. 2011CBA00205, No. 2009CB929504, and No. 2011CB922104), by NSFC (Grants No. 11074111, No. 11174125, and No. 11023002), by PAPD of Jiangsu Higher Education Institutions, by NCET,  by the GRF (HKU7058/11P) and CRF (HKU-8/11G) of the RGC of Hong Kong. 
\end{acknowledgments}

\end{document}